\documentclass[preprint]{emulateapj}
\usepackage{color}

\slugcomment{}

\shorttitle{The first LBGs individually detected in the FIR by
\textit{Herschel}}
\shortauthors{Burgarella et al.}

\begin{document}


\title{HerMES: Lyman Break Galaxies individually detected at 
\mbox{\boldmath{$0.7\leq z \leq 2.0$}} in GOODS-N with \textit{Herschel}/SPIRE}


\author{D.~Burgarella\altaffilmark{1}}
\affil{Laboratoire d'Astrophysique de Marseille, OAMP, Universit\'e Aix-Marseille, CNRS, 38 rue Fr\'ed\'eric Joliot-Curie, 13388 Marseille cedex 13, France}
\email{denis.burgarella@oamp.fr}
\author{S.~Heinis\altaffilmark{1}}
\affil{Laboratoire d'Astrophysique de Marseille, OAMP, Universit\'e Aix-Marseille, CNRS, 38 rue Fr\'ed\'eric Joliot-Curie, 13388 Marseille cedex 13, France}
\email{sebastien.heinis@oamp.fr}
\author{G.~Magdis\altaffilmark{2}}
\affil{Laboratoire AIM-Paris-Saclay, CEA/DSM/Irfu - CNRS - Universit\'e Paris Diderot, CE-Saclay, pt courrier 131, F-91191 Gif-sur-Yvette, France}
\author{R.~Auld\altaffilmark{3}}
\affil{Cardiff School of Physics and Astronomy, Cardiff University, Queens
 Buildings, The Parade, Cardiff CF24 3AA, UK}
\author{A.~Blain\altaffilmark{13}}
\affil{Institute for Astronomy, University of Edinburgh, Royal Observatory,
 Blackford Hill, Edinburgh EH9 3HJ, UK}
\author{J.~Bock\altaffilmark{4,5}}
\affil{California Institute of Technology, 1200 E. California Blvd., Pasadena,
 CA 91125, USA and Jet Propulsion Laboratory, 4800 Oak Grove Drive, Pasadena, CA 91109, USA}
\author{D.~Brisbin\altaffilmark{6}}
\affil{Space Science Building, Cornell University, Ithaca, NY, 14853-6801, USA}
\author{V.~Buat \altaffilmark{1}}
\affil{Laboratoire d'Astrophysique de Marseille, OAMP, Universit\'e Aix-Marseille, CNRS, 38 rue Fr\'ed\'eric Joliot-Curie, 13388 Marseille cedex 13, France}
\author{P.~Chanial\altaffilmark{7}}
\affil{Astrophysics Group, Imperial College London, Blackett Laboratory,
 Prince Consort Road, London SW7 2AZ, UK}
\author{D.L.~Clements\altaffilmark{7}}
\affil{Astrophysics Group, Imperial College London, Blackett Laboratory,
 Prince Consort Road, London SW7 2AZ, UK}
\author{A.~Cooray\altaffilmark{8,4}}
\affil{Dept. of Physics \& Astronomy, University of California, Irvine,
 CA 92697, USA and California Institute of Technology, 1200 E. California Blvd., Pasadena,
 CA 91125, USA}
\author{S.~Eales\altaffilmark{3}}
\affil{Cardiff School of Physics and Astronomy, Cardiff University, Queens
 Buildings, The Parade, Cardiff CF24 3AA, UK}
\author{A.~Franceschini\altaffilmark{9}}
\affil{Dipartimento di Astronomia, Universit\`{a} di Padova, vicolo
 Osservatorio, 3, 35122 Padova, Italy}
\author{E.~Giovannoli\altaffilmark{1}}
\affil{Laboratoire d'Astrophysique de Marseille, OAMP, Universit\'e Aix-Marseille, CNRS, 38 rue Fr\'ed\'eric Joliot-Curie, 13388 Marseille cedex 13, France}
\author{J.~Glenn\altaffilmark{10}}
\affil{Dept. of Astrophysical and Planetary Sciences, CASA 389-UCB,
 University of Colorado, Boulder, CO 80309, USA}
\author{E.A.~Gonz\'alez~Solares\altaffilmark{11}}
\affil{Institute of Astronomy, University of Cambridge, Madingley Road,
 Cambridge CB3 0HA, UK}
\author{M.~Griffin\altaffilmark{3}}
\affil{Cardiff School of Physics and Astronomy, Cardiff University, Queens
 Buildings, The Parade, Cardiff CF24 3AA, UK}
\author{H.S.~Hwang\altaffilmark{2}}
\affil{Laboratoire AIM-Paris-Saclay, CEA/DSM/Irfu - CNRS - Universit\'e Paris Diderot, CE-Saclay, pt courrier 131, F-91191 Gif-sur-Yvette, France}
\author{O.~Ilbert\altaffilmark{1}}
\affil{Laboratoire d'Astrophysique de Marseille, OAMP, Universit\'e Aix-Marseille, CNRS, 38 rue Fr\'ed\'eric Joliot-Curie, 13388 Marseille cedex 13, France}
\author{L.~Marchetti\altaffilmark{9}}
\affil{Dipartimento di Astronomia, Universit\`{a} di Padova, vicolo
 Osservatorio, 3, 35122 Padova, Italy}
\author{A.M.J.~Mortier\altaffilmark{7}}
\affil{Astrophysics Group, Imperial College London, Blackett Laboratory,
 Prince Consort Road, London SW7 2AZ, UK}
\author{S.J.~Oliver\altaffilmark{14}}
\affil{Astronomy Centre, Dept. of Physics \& Astronomy, University of Sussex,
 Brighton BN1 9QH, UK}
\author{M.J.~Page\altaffilmark{15}}
\affil{Mullard Space Science Laboratory, University College London, Holmbury
 St.\ Mary, Dorking, Surrey RH5 6NT, UK}
\author{A.~Papageorgiou\altaffilmark{3}}
\affil{Cardiff School of Physics and Astronomy, Cardiff University, Queens
 Buildings, The Parade, Cardiff CF24 3AA, UK}
\author{C.P.~Pearson\altaffilmark{16,17}}
\affil{RAL Space, Rutherford Appleton Laboratory, Chilton, Didcot, Oxfordshire
OX11 0QX, UK  and Institute for Space Imaging Science, University of Lethbridge,
 Lethbridge, Alberta, T1K 3M4, Canada}
\author{I.~P{\'e}rez-Fournon\altaffilmark{18,19}}
\affil{Instituto de Astrof{\'\i}sica de Canarias (IAC), E-38200 La Laguna,
 Tenerife, Spain and Departamento de Astrof{\'\i}sica, Universidad de La Laguna (ULL),
 E-38205 La Laguna, Tenerife, Spain}
\author{M.~Pohlen\altaffilmark{3}}
\affil{Cardiff School of Physics and Astronomy, Cardiff University, Queens
 Buildings, The Parade, Cardiff CF24 3AA, UK}
\author{J.I.~Rawlings\altaffilmark{15}}
\affil{Mullard Space Science Laboratory, University College London, Holmbury
 St.\ Mary, Dorking, Surrey RH5 6NT, UK}
\author{G.~Raymond\altaffilmark{3}}
\affil{Cardiff School of Physics and Astronomy, Cardiff University, Queens
 Buildings, The Parade, Cardiff CF24 3AA, UK}
\author{D.~Rigopoulou\altaffilmark{16,20}}
\affil{Space Science \& Technology Department, Rutherford Appleton Laboratory,
 Chilton, Didcot, Oxfordshire OX11 0QX, UK and Astrophysics, Oxford University, Keble Road, Oxford OX1 3RH, UK}
\author{G.~Rodighiero\altaffilmark{9}}
\affil{Dipartimento di Astronomia, Universit\`{a} di Padova, vicolo
 Osservatorio, 3, 35122 Padova, Italy}
\author{I.G.~Roseboom\altaffilmark{14}}
\affil{Astronomy Centre, Dept. of Physics \& Astronomy, University of Sussex,
 Brighton BN1 9QH, UK}
\author{M.~Rowan-Robinson\altaffilmark{7}}
\affil{Astrophysics Group, Imperial College London, Blackett Laboratory,
 Prince Consort Road, London SW7 2AZ, UK}
\author{Douglas~Scott\altaffilmark{21}}
\affil{Department of Physics \& Astronomy, University of British Columbia,
 6224 Agricultural Road, Vancouver, BC V6T~1Z1, Canada}
\author{N.~Seymour\altaffilmark{15}}
\affil{Mullard Space Science Laboratory, University College London, Holmbury
 St.\ Mary, Dorking, Surrey RH5 6NT, UK}
\author{A.J.~Smith\altaffilmark{14}}
\affil{Astronomy Centre, Dept. of Physics \& Astronomy, University of Sussex,
 Brighton BN1 9QH, UK}
\author{M.~Symeonidis\altaffilmark{15}}
\affil{Mullard Space Science Laboratory, University College London, Holmbury
 St.\ Mary, Dorking, Surrey RH5 6NT, UK}
\author{K.E.~Tugwell,\altaffilmark{15}}
\affil{Mullard Space Science Laboratory, University College London, Holmbury
 St.\ Mary, Dorking, Surrey RH5 6NT, UK}
\author{M.~Vaccari\altaffilmark{9}}
\affil{Dipartimento di Astronomia, Universit\`{a} di Padova, vicolo
 Osservatorio, 3, 35122 Padova, Italy}
\author{J.D.~Vieira\altaffilmark{4}}
\affil{California Institute of Technology, 1200 E. California Blvd., Pasadena,
 CA 91125, USA}
\author{M.~Viero\altaffilmark{4}}
\affil{California Institute of Technology, 1200 E. California Blvd., Pasadena,
 CA 91125, USA}
\author{L.~Vigroux\altaffilmark{22}}
\affil{Institut d'Astrophysique de Paris, UMR 7095, CNRS, UPMC Univ. Paris 06,
 98bis boulevard Arago, F-75014 Paris, France}
\author{L.~Wang\altaffilmark{14}}
\affil{Astronomy Centre, Dept. of Physics \& Astronomy, University of Sussex,
 Brighton BN1 9QH, UK}

\and
\author{G.~Wright\altaffilmark{12}}
\affil{UK Astronomy Technology Centre, Royal Observatory, Blackford Hill,
 Edinburgh EH9 3HJ, UK}



\begin{abstract}
As part of the {\it Herschel\/} Multi-tiered Extragalactic Survey we have
investigated the rest-frame far-infrared (FIR) properties of a sample of more
than 4800 Lyman Break Galaxies (LBGs) in the Great Observatories Origins
Deep Survey North field.  Most
LBGs are not detected individually, but we do detect a sub-sample of 12
objects at $0.7\,{<}\,z\,{<}\,1.6$ and one object at $z\,{=}\,2.0$.
The ones detected by {\it Herschel\/} SPIRE have
redder observed NUV-U and U-R colors than the others, while the undetected ones have colors consistent
with average LBGs at $z\,{>}\,2.5$.  The UV-to-FIR spectral energy distributions of the objects
detected in the rest-frame FIR are investigated using the code {\sc cigale} to
estimate physical parameters. We find that LBGs detected by SPIRE are high mass, luminous infrared galaxies.
It appears that LBGs are located in a triangle-shaped region in the $A_{\rm FUV}$ 
vs. $Log L_{\rm FUV}$ diagram limited by $A_{\rm FUV}=0$ at the bottom and by 
a diagonal following the temporal evolution of the most massive galaxies from the bottom-right to the top-left of the diagram. This upper envelop can be used as upper limits for the UV dust attenuation as a function of L$_{\rm FUV}$. The limits of this region are well explained using a closed-box model, where the chemical evolution of galaxies produces metals, which in turn lead to higher dust attenuation when the galaxies age.
\end{abstract}




%% Note that for sources with brackets in their names, e.g., [WEG2004] 14h-090,

\section{Introduction}

%Dropout galaxies are selected from a color-color diagram \citep{steidel96} that highlights the presence of the Lyman break in the far-ultraviolet (FUV).
%This Lyman break technique selects star-forming galaxies at all redshifts.
%Depending on the redshift that one desires to explore, a different set of filters is used to detect the Lyman break feature. 

%The safest way to estimate the total star formation rates (SFRs) of galaxies is to consider the energy budget involving far-ultraviolet (FUV) and far-infrared (FIR) measurements (e.g., \citealt{buat96}). This is performed via the ratio of the dust emission measured across the IR ($L_{\rm IR}=L [8{-}1000\,\mu{\rm m]}$) to the FUV emission $L_{\rm IR}/L_{\rm FUV}$ (e.g., \citealt{burgarella05}).  
%On the other hand, based on a sample of local starbursts, \citet{meurer99} suggest that the slope of the UV continuum $\beta$ (defined as $ f_\lambda \propto \lambda^\beta$) could be used to estimate the amount of dust attenuation. 
%While still debated in detail, it is still reasonable to use the relation between $\beta$ and $L_{\rm IR}/L_{\rm FUV}$ to correct for the dust attenuation of UV-selected galaxies.

The safest way to estimate the total star formation rates (SFR) of galaxies is to consider the energy budget involving far-ultraviolet (FUV) and far-infrared (FIR) measurements (e.g., \citealt{buat96}). But, because only a small number of individual Lyman Break Galaxies (LBGs) have been
detected in the FIR/submm range \citep[e.g.,][]{chapman00, chapman09, siana09}, we need to observe this type of  galaxies at lower redshifts to understand their dust emission. \cite{burgarella07} detected dropout galaxies at $z \sim 1$ at $24\,\mu$m with {\it Spitzer}.
But, the dust luminosities estimated from the rest frame $12\,\mu$m
flux density is far from the peak of the dust emission and could provide biased SFR estimates.

We observe in the FIR a sample of LBGs at $0.7\,{<}\,z\,{<}\,1.6$ (FUV dropouts) and at
$1.6\,{<}\,z\,{<}\,2.8$ (near-UV or NUV dropouts). We use the SPIRE instrument \citep{griffin10} on
{\it Herschel\/} \citep{pilbratt10} with observations from HerMES
\citep{oliver10,oliver10b}\footnote{http://hermes.sussex.ac.uk}. This is the
first opportunity to estimate directly the dust luminosity (or upper limits)
of unlensed LBGs from FIR data. We assume $\Omega_{\rm m} = 0.3$,
$\Omega_{\Lambda} = 0.7$, and
 $H_0 = 71\,{\rm km}\,{\rm s}^{-1}\,{\rm Mpc}^{-1}$ and use AB magnitudes
throughout.  

\section{Data}

%% In a manner similar to \objectname authors can provide links to dataset
%% hosted at participating data centers via the \dataset{} command.  The
%% second curly bracket argument is printed in the text while the first
%% parentheses argument serves as the valid data set identifier.  Large
%% lists of data set are best provided in a table (see Table 3 for an example).
%% Valid data set identifiers should be obtained from the data center that
%% is currently hosting the data.
%%
%% Note that AASTeX interprets everything between the curly braces in the 
%% macro as regular text, so any special characters, e.g., "#" or "_," must be 
%% preceded by a backslash. Otherwise, you will get a LaTeX error when you 
%% compile your manuscript.  Special characters do not 
%% need to be escaped in the optional, square-bracket argument.

\subsection{LBG samples}

\begin{figure*}
  \begin{center}$
  \begin{array}{cc}
   \includegraphics [width=140mm, height=180mm, angle=0] {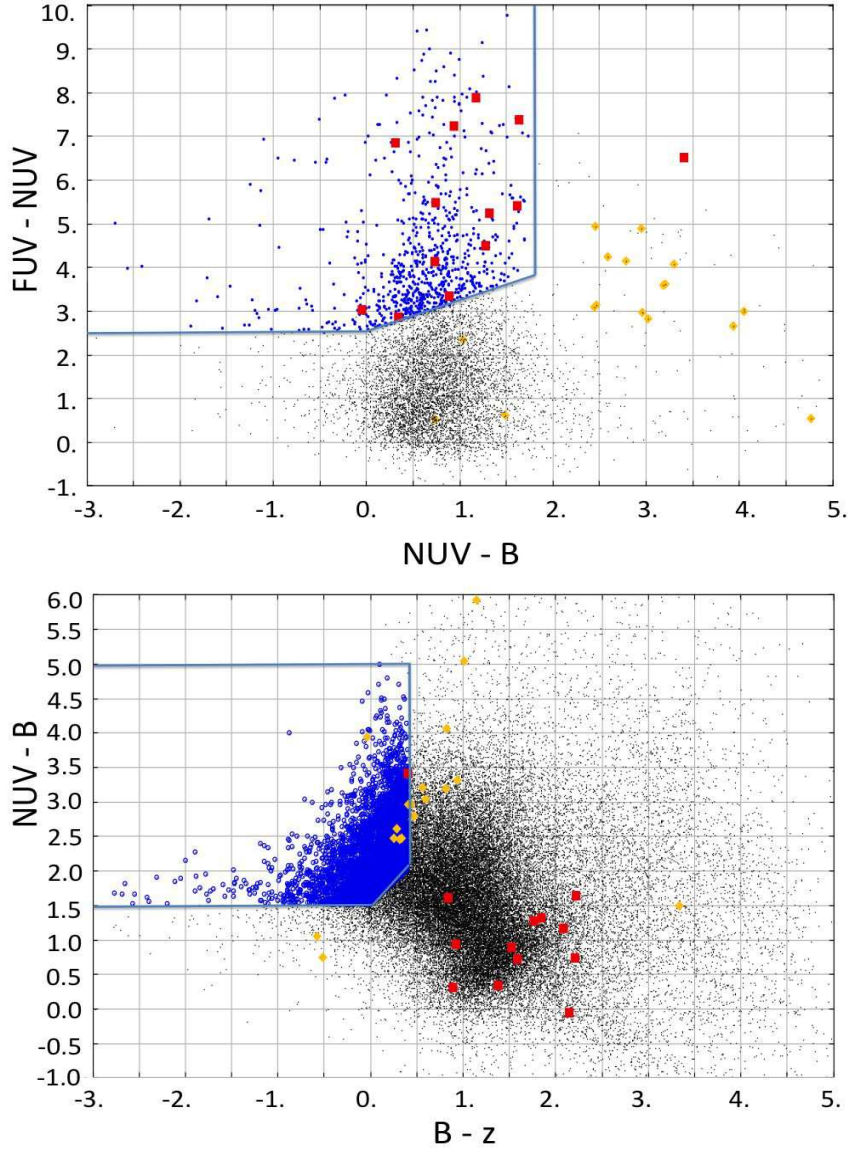} 
 \end{array}$
 \end{center}
   \caption{The upper panel shows the selection for $0.7\,{<}\,z\,{<}\,1.6$
dropouts: $m[{\rm FUV}]-m[{\rm NUV}] \geq 0.7 (m[{\rm NUV}] - B) + 2.5$;
$m[{\rm FUV}]-m[{\rm NUV}] \geq 2.5$; $m[{\rm NUV}]-B \leq 1.8$; and
${\rm SNR(NUV)} \geq 3$.  This corresponds to 740 objects, with an average
${\rm SNR(NUV)} = 8.6$.  The lower panel shows the selection for $1.6<z<2.8$
dropouts: $m[{\rm NUV}] - B \geq 1.25 (B - z) + 1.5$;
$m[{\rm NUV}] - B \geq 1.5$; $m[{\rm NUV}] - B \leq 5.0$; $B - z \leq 0.4$;
and ${\rm SNR}(B) \geq 3.0$, which corresponds to 4107 objects, with a
sample average ${\rm SNR(}B{\rm )} = 9.7$. The LBGs are shown as blue dots
and the SPIRE-detected LBGs by big (red) boxes. Both LBG samples are plotted
on each panel. Yellow diamonds are stars.}
    \label{selLBGs}%
 \end{figure*}

Observations of the Great Observatories Origins Deep Survey North (GOODS-N)
were secured as part of the {\it Herschel\/}
Multi-Tiered Extragalactic Survey \citep[HerMES,][]{oliver10} by {\it GALEX\/} in FUV and NUV. We define two samples of
galaxies in two redshift ranges corresponding to FUV dropouts and NUV dropouts
(Fig.~\ref{selLBGs}).

The photometry is performed with IRAF {\sc daophot~ii} \citep{stetson87}
in the NUV, and in the FUV with the NUV coordinates. Using {\sc addstar}, 
the completeness is estimated to 80\%
down to $m[{\rm FUV}]=24.9$ and $m[{\rm NUV}]=24.2$. We use {\sc cigale}
\citep{noll09} to build models in the range $0\,{\le}\,z\,{\le}\,3$ and delimit
the regions corresponding to LBGs at $0.7\,{<}\,z\,{<}\,1.6$ and
$1.6\,{<}\,z\,{<}\,2.8$ in the color-color diagrams.\footnote{Note that the
code accounts for IGM attenuation.}
The sample is cross-correlated with the
$R$-selected \cite{capak04} multiwavelength catalog over $0.4\,{\rm deg}^2$,
with a search radius of 1\arcsec, producing a catalog of 86{,}768 entries
(46{,}076 with {\it GALEX\/} data, 47{,}450 with optical data and 6{,}979
objects with both). The $U$-band data were collected using the Kitt Peak
National Observatory 4-m telescope. The $B$-, $V$-, $R$-, $I$-, and $z$-band
data were collected using the Subaru 8.2-m telescope and Suprime-Cam instrument
with 5$\sigma$ limiting magnitudes of $U = 27.1$, $B=26.9$, $V = 26.8$,
$R = 26.6$, $I = 25.6$ and $z = 25.4$. Photometric redshifts are computed using
{\sc Le Phare} \citep{arnouts02, ilbert09} and we use them in addition to spectroscopic
redshifts from \cite{barger08}. In the redshift range $0.7\,{<}\,z\,{<}\,1.6$, we compare photometric to spectroscopic redshifts (784 galaxies) and find $\sigma_z$ / (1+z) = 0.036 while we have (24 galaxies) $\sigma_z$ / (1+z) = 0.125 at $1.6\,{<}\,z\,{<}\,2.8$.

From the lower redshift catalog, 27 objects (4\%) have photometric redshifts
at $z\,{<}\,0.7$, 696 objects (94\%) are in our redshift range of
$0.7\,{<}\,z\,{<}\,1.6$ and 17 objects (2.3\%) are at $z\,{>}\,1.6$.
For the higher redshift catalog, 223 objects (5\%) have photometric redshifts
at $z\,{<}\,1.6$, 3859 objects (94\%) are in our redshift range of
$1.6\,{<}\,z\,{<}\,2.8$, and 25 objects (0.6\%) are at $z\,{>}\,2.8$.

\subsection{Matching with SPIRE}

We cross-correlate the LBG samples with the HerMES multi-wavelength catalog
from \citet{roseboom10}. It uses 24$\,\micron$ sources as
a prior, so we restrict ourselves to the deep GOODS-N MIPS region of
$19\arcmin\times12\arcmin$, i.e., $0.063\,{\rm deg}^2$.  Within this region we
have 260 low-$z$ LBGs and 1558 high-$z$. This HerMES catalog contains 1951
24$\,\mu$m sources as an input to the association process. A search radius of
1\arcsec\ is chosen, since the 3.6$\,\micron$ positions are good to that
accuracy.  Given the source density of the HerMES catalog this should return
$<1$\% spurious matches. We find 86 matches between this list and our LBG
samples, with 63 matches to the low-$z$ sample and 23 matches to the high-$z$
sample, with $\sim 2$ (3.2\%) and $\sim 11$ (47.8\%), respectively, expected by
chance.

The HerMES catalog gives SPIRE measurements for all MIPS sources.  To define
confident SPIRE detections, we require flux densities higher than 7.6, 9.2 or
10.4\,mJy
at 250, 350 and 500$\,\mu$m respectively.  This flux limit corresponds to
$2 \times \sigma_{\rm faint}$, the clipped map confusion
noise, where $\sigma_{\rm faint}$ is estimated by \cite{nguyen10} from the map
variance after removing pixels brighter than $5\sigma_{\rm conf}$ (the raw
map confusion noise). Imposing these more reasonable flux density limits gives
14 low-$z$ candidates and one high-$z$ candidate.

%We then checked in the SPIRE maps that there are no bright objects close to our candidates by examining objects in the UV to FIR images. 
We define SPIRE detections (listed in
Table~\ref{Table}) whenever the SNR is larger than 3 and their ``purity''
index \citep[see][]{roseboom10,brisbin2010} is larger than 0.2.  Note that
according to this definition there are no detections at $500\,\mu$m.
The ``purity'' of the SPIRE flux density is
estimated from the ratio of this source's 24$\,\mu$m flux density to the
24$\,\mu$m one smoothed with the SPIRE beam at this position. For no
``pollution'', the purity is 1 and decreases when there are possible
contributions by other sources.  We additionally inspected the images around
each candidate to check for contamination by neighbors.  Based on the purity index, 
we exclude 2 objects at $0.7\,{<}\,z\,{<}\,1.6$.
All but one of the sources on the final sample (actually the highest redshift
LBG) have SNR$\ga$ 6 (see Table~\ref{Table}). 

The final $0.7\,{<}\,z\,{<}\,1.6$ HerMES LBG sample contains 12 objects and the
$1.6\,{<}\,z\,{<}\,2.8$ sample contains 1 object. Although it is difficult to
determine what fraction of $z\sim$2 LBGs should be detected in the FIR,
we can use the
{\it Spitzer\/} $24\,\mu$m data to estimate how many $z\sim1$ LBGs are
expected to be detected at 250$\,\mu$m. From the present sample, the mean ratio
$S_{\rm 250}/S_{24} = 54 \pm 11$ for the SPIRE-detected LBGs and so, the
present 250$\,\mu$m detection limit of 7.6\,mJy correspond to 
140$\,\mu$Jy at 24$\,\mu$m. At this level, and based on the 24$\,\mu$m catalog, 
we would expect 20 $z\sim1$ LBGs to be detected with SPIRE at 250$\,\mu$m.
We find 12 objects, i.e., 60\% of the
expectation. This is consistent with the completeness at this flux level in
\citet{roseboom10}.

\subsection{SED fitting}

Dust luminosities ($L_{\rm IR}=L [8{-}1000\,\mu{\rm m]}$) and other parameters are estimated
using {\sc cigale} (\citealt{noll09})\footnote{http://www.oamp.fr/cigale}. {\sc cigale} performs a
Bayesian analysis to estimate parameter by fitting models to the
UV-to-submm SEDs. One can select among two single stellar population libraries
and several IR models/templates. An AGN component can also be added to estimate
the AGN fraction (contribution by a potential AGN to $L_{\rm IR}$).
The parameters of the dust attenuation law can be modified and {\sc cigale}
allows for two separate stellar populations with a multiphase dust treatment.

We use the VLA 1.4\,GHz radio data from \cite{morrison10} which provides data down to a RMS noise of 3.9 $\mu$Jy per beam. The stellar emission is based on
\cite{maraston05}, while the dust emission is based on \cite{dale02} templates.
{\sc cigale} provides dust luminosities $L_{\rm IR}$, while FUV luminosities
$L_{\rm FUV}$ are derived at $\lambda_{\rm rest}=153\,$nm and are defined
through the quantity
$\nu L_{\nu}$.  An energy budget is performed during the fit, and the
maximum value allowed for $L_{\rm IR}$ has to be consistent with the energy
moved by dust grains from the UV-optical range to the FIR range.

\begin{figure*}
 \begin{center}
   \includegraphics [width=160mm, height=200mm, angle=0] {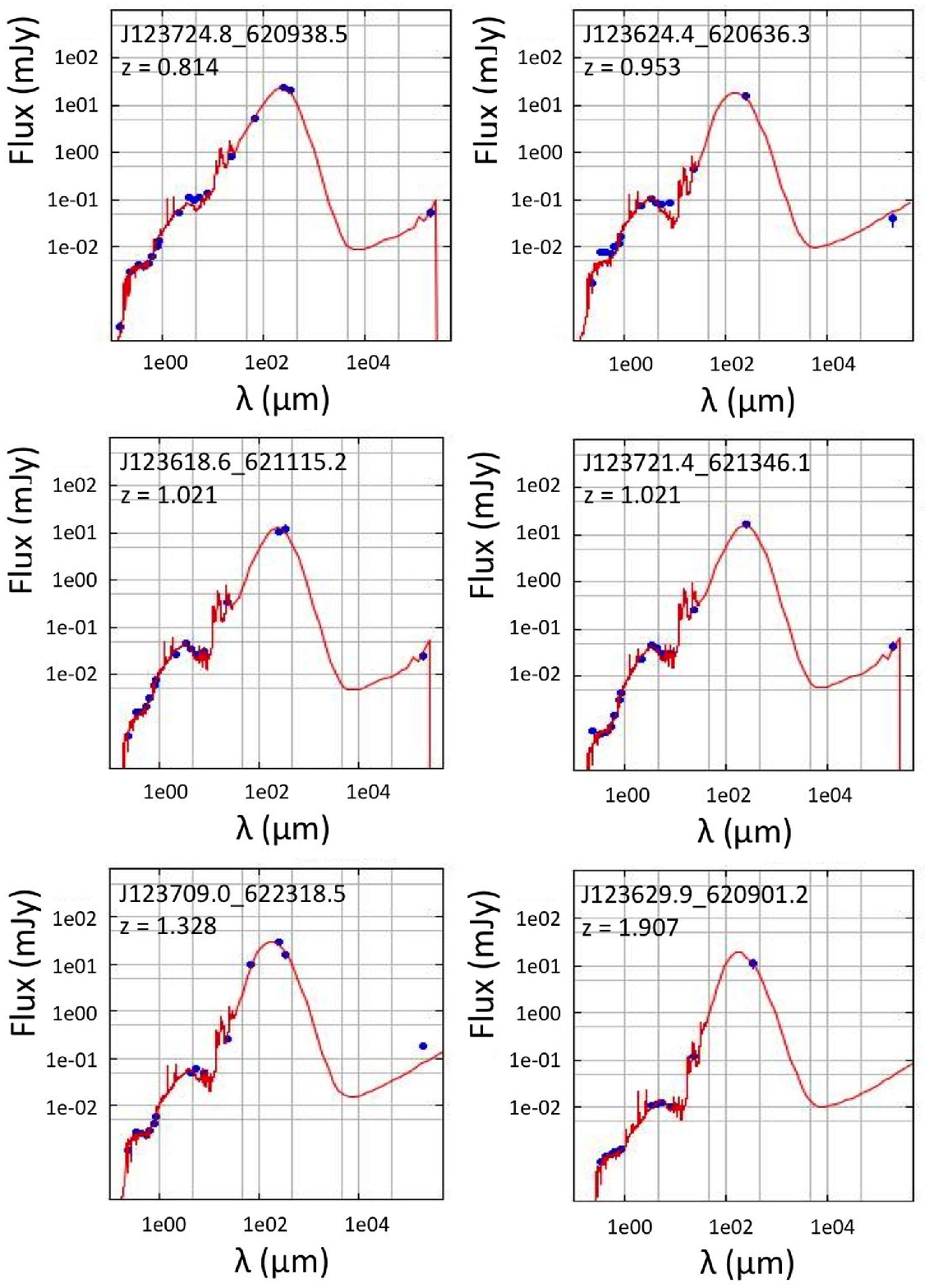}
 \end{center}
   \caption{Observed spectral energy distributions of LBGs superimposed on
best fit models ($S_\nu[{\rm mJy}]$ vs. $\log \lambda[\mu$m]).
Note that J123633.2$+$620834.9 very likely hosts an AGN, as suggested by
the {\sc cigale} SED fitting. }
    \label{sedlbg}%%
 \end{figure*}

Individual SEDs with the best models selected by {\sc cigale} are shown in
Fig.~\ref{sedlbg} for the five LBGs with radio data and for the 
($z=1.9$) LBG.  Observational and 
physical parameters are given in Table~\ref{Table}.
Fig.~\ref{sedlbg} shows that {\sc cigale} is able to fit the SEDs from the
FUV to the radio successfully. The average FUV luminosity of the SPIRE
detected objects is
$ \log\langle(L_{\rm FUV}/{\rm L}_\odot)\rangle = (10.7 \pm 0.2)$,
and most of them are therefore UV-Luminous Galaxies.
Their average dust luminosity is
$ \log\langle(L_{\rm IR}/{\rm L}_\odot) \rangle= (11.9 \pm 0.1)$
and their stellar average mass is
$ \log\langle(M_*/{\rm M}_\odot) \rangle = (11.0 \pm 0.5)$.
A comparison with \cite{magdis10} shows that the average stellar mass is
similar to the average stellar mass
($ \log\langle(M_*/{\rm M}_\odot) \rangle = 11$) of $z \sim 3$ LBGs detected
at $\lambda = 8\,\mu$m. The dust temperatures $T_{\rm d}$ are estimated for
a few objects by fitting modified black bodies with an emissivity index of 1.5.
We find that two of the low-$z$ LBGs have $T_{\rm d}$ $\sim24\,$K, while the
high-$z$ ULIRG has $T_{\rm d} = 52\,$K (see Table~\ref{Table}). The two 
low-$z$ LBGs have quite low temperatures
compared with ULIRGS/SMGs, but may be typical of star-forming galaxies in
general, while the high-$z$ one is similar to e.g. Arp220.

 \begin{table*}
 \caption{Observational data and physical parameters deduced by {\sc cigale}. The column origin redshift is "specz" if the redshift is spectroscopic or the 3$\sigma$ uncertainty otherwise.}
 \label{Table}
% \scalebox{0.45}{%
 \scalebox{0.65}{%
 \begin{tabular}{@{}lcccccccccccccc}
  \hline \hline
ID  & $z$ & origin & $m[{\rm NUV}]$ & $U$ & $S_{24}$  & $S_{250}$  & $S_{350}$ & $S_{1.4GHz}$ &
 $\log M_*$ & SFR & $\log L_{\rm IR}$ & $A_{\rm FUV}$ & $\log L_{\rm FUV}$ &
 $T_{\rm d}$\\
& & redshift & [AB] & [AB] & [$\mu$Jy] & [mJy] & [mJy] & [mJy] & [M$_\odot$] &
 [M$_\odot\,{\rm yr}^{-1}$]  &[M$_\odot]$ &  [mag] & [L$_\odot$] & [K] \\
  \hline
J123624.6$+$620610.2 & 0.75 & 0.11 & 22.87 $\pm$ 0.04 & 23.25 $\pm$ 0.04 &
 93.2 $\pm$ 8.3 & 8.0 $\pm$ 0.9 & --- &  ---  & 10.0 $\pm$ 0.1 & 23.5 $\pm$ 0.048 &
 11.1 $\pm$ 0.039 & 1.96 $\pm$ 0.14 & 10.4 $\pm$ 0.6 & --- \\
J123547.4$+$621005.9 & 0.81 & 0.24 & 24.07 $\pm$ 0.07 & 25.33 $\pm$ 0.24 &
 86.3 $\pm$ 7.6 & 11.3 $\pm$ 0.9 & 12.1 $\pm$ 1.5 &  --- & 10.2 $\pm$ 0.1 &
 34.8 $\pm$ 0.030 & 11.3 $\pm$ 0.007 & 3.68 $\pm$ 0.28 & 9.9 $\pm$ 1.4 &
 --- \\
J123724.8$+$620938.5 & 0.81 & 0.13 & 22.80 $\pm$ 0.04 & 23.59 $\pm$ 0.06 &
 249.0 $\pm$ 6.7 & 10.3 $\pm$ 0.8 & --- &  --- & 10.7 $\pm$ 0.1 & 39.4 $\pm$ 0.090 &
 11.3 $\pm$ 0.094 & 2.42 $\pm$ 0.28 & 10.5 $\pm$ 0.9 & --- \\
J123633.2$+$620834.9 & 0.93 & specz & 22.41 $\pm$ 0.03 & 22.84 $\pm$ 0.03 &
 779.0 $\pm$ 7.9 & 22.4 $\pm$ 1.2 & 19.2 $\pm$ 1.8 & 50.8 $\pm$ 10.2 & 11.1 $\pm$ 0.2 &
 119.7 $\pm$ 0.040 & 11.9 $\pm$ 0.027 & 2.84 $\pm$ 0.18 & 10.9 $\pm$ 1.1 &
 24.3 \\
J123714.4$+$622112.3 & 0.94 & specz & 24.20 $\pm$ 0.08 & 25.17 $\pm$ 0.21 &
 212.0 $\pm$ 4.6 & 9.2 $\pm$ 1.2 & 12.0 $\pm$ 2.7 &  --- & 10.7 $\pm$ 0.2 & 
 43.9 $\pm$ 0.059 & 11.4 $\pm$ 0.050 & 3.46 $\pm$ 0.32 & 10.1 $\pm$ 1.3 &
 --- \\
J123624.4$+$620836.3 & 0.95 & specz & 23.22 $\pm$ 0.05 & 23.83 $\pm$ 0.07 &
 695.0 $\pm$ 11.5 & 21.1 $\pm$ 2.8 & 22.9 $\pm$ 7.0 & ---  & 11.3 $\pm$ 0.2 &
 116.4 $\pm$ 0.182 & 11.9 $\pm$ 0.156 & 3.45 $\pm$ 0.50 & 10.6 $\pm$ 1.3 &
 24.2 \\
J123614.4$+$620718.5 & 0.97 & specz & 23.88 $\pm$ 0.07 & 25.32 $\pm$ 0.23 &
 386.0 $\pm$ 8.7 & 9.2 $\pm$ 1.0 & 10.2 $\pm$ 1.2 &  --- & 11.0 $\pm$ 0.2 &
 61.8 $\pm$ 0.275 & 11.7 $\pm$ 0.218 & 3.52 $\pm$ 0.78 & 10.3 $\pm$ 1.4 &
 --- \\
J123721.4$+$621346.1 & 1.02 & specz & 24.53 $\pm$ 0.10 & 24.41 $\pm$ 0.11 &
 235.0 $\pm$ 7.5 & 15.4 $\pm$ 2.6 & --- & 41.6 $\pm$ 8.7 & 11.0 $\pm$ 0.2 & 113.2 $\pm$ 0.134 &
 11.9 $\pm$ 0.120 & 4.52 $\pm$ 0.38 & 10.2 $\pm$ 1.7 & --- \\
J123618.6$+$621115.2 & 1.02 & specz & 21.79 $\pm$ 0.03 & 23.40 $\pm$ 0.05 &
 404.0 $\pm$ 5.4 & 14.5 $\pm$ 2.3 & --- & 36.5 $\pm$ 11.1 & 11.3 $\pm$ 0.2 &
 135.5 $\pm$ 0.161 & 11.9 $\pm$ 0.133 & 2.60 $\pm$ 0.42 & 11.0 $\pm$ 0.9 &
 --- \\
J123722.5$+$621356.6 & 1.02 & specz & 23.46 $\pm$ 0.06 & 24.78 $\pm$ 0.15 &
 317.0 $\pm$ 6.2 & 9.7 $\pm$ 1.2 & 11.5  $\pm$ 2.3  & 23.7 $\pm$ 4.7 & 10.9 $\pm$ 0.2 &
 74.0 $\pm$ 0.155 & 11.7 $\pm$ 0.145 & 3.23 $\pm$ 0.43 & 10.4 $\pm$ 1.2 &
 --- \\
J123808.9$+$621847.5 & 1.04 & 0.29 & 22.89 $\pm$ 0.06 & 23.66 $\pm$ 0.06 &
 218.0 $\pm$ 8.8 & 9.8 $\pm$ 1.1 & --- &  --- & 10.8 $\pm$ 0.3 & 78.2 $\pm$ 0.168 &
 11.6 $\pm$ 0.156 & 2.46 $\pm$ 0.40 & 10.8 $\pm$ 0.8 & --- \\
J123709.0$+$622318.5 & 1.33 & 0.28 & 22.91 $\pm$ 0.05 & 23.91 $\pm$ 0.08 &
 247.0 $\pm$ 6.7 & 28.2 $\pm$ 1.3 & 14.3 $\pm$ 2.4 & 176.4 $\pm$ 13.9 & 11.2 $\pm$ 0.1 &
 448.8 $\pm$ 0.066 & 12.4 $\pm$ 0.027 & 3.82 $\pm$ 0.18 & 11.1 $\pm$ 1.3 &
 51.8 \\
J123629.6$+$620901.2 & 1.97 & 0.67 & 24.33 $\pm$ 0.09 & --- & 110.0 $\pm$ 6.0 &
 --- & 10.5 $\pm$ 2.4 &  --- & 10.5 $\pm$ 0.1 & 893.3 $\pm$ 0.036 & 12.5 $\pm$ 0.035 &
 4.30 $\pm$ 0.20 & 11.1 $\pm$ 1.5 & --- \\
  \hline
 \end{tabular}}
\end{table*}

%% In this section, we use  the \subsection command to set off
%% a subsection.  \footnote is used to insert a footnote to the text.


%% This section also includes several of the displayed math environments
%% mentioned in the Author Guide.

\section{Dust attenuation of Lyman Break Galaxies}

\begin{figure*}
   \includegraphics [width=160mm, angle=0] {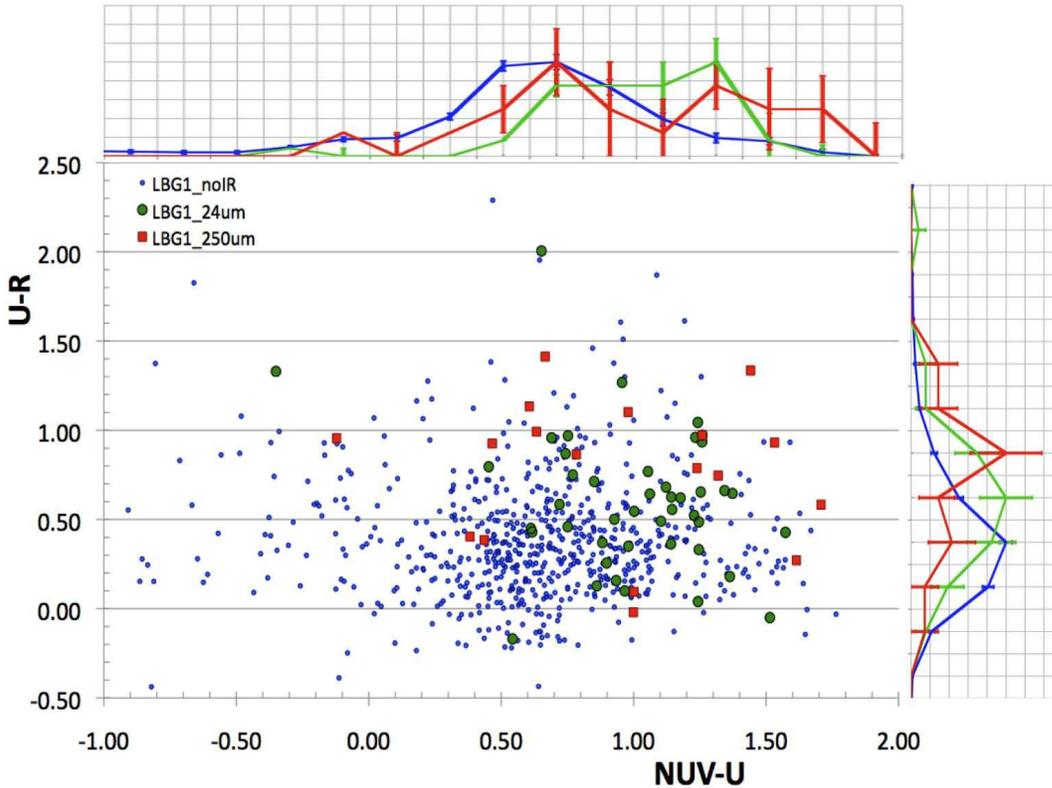} 
   \caption{In this $(U - R)$ vs. $(m[{\rm NUV}] - U)$ diagram, blue points
are LBGs which are undetected at 24$\,\mu$m, green dots are those detected
at 24$\,\mu$m and red boxes are those detected in the FIR with {\it Herschel}.
We see on the marginal distributions that LBGs get redder with increasing maximum wavelength of detection,
especially in the U-R color. This is expected if IR-bright LBGs are more attenuated 
than IR-faint LBGs.}
    \label{color}%
 \end{figure*}

Our LBG SEDs in the rest-frame UV are in very good agreement with the rest-frame UV spectra of the \cite{shapley03}'s
composite spectrum at $z\,{\sim}\,3$.
Whatever the redshift range, LBGs seem to present the same starburst
characteristics in the rest-frame UV, which is expected,
given that similar rest-frame color selection criteria are used.
However, the colors become redder when the objects are detected at
24$\,\mu$m by {\it Spitzer\/} and even redder when they are detected at
250$\,\mu$m by {\it Herschel\/} SPIRE. This can be interpreted as being
due to higher dust attenuations (see Fig.~\ref{color}), and is consistent
with \cite{burgarella07}, who found that both{\it Spitzer}-detected and
undetected LBGs have about the same stellar population ages, but the latter
are more extinguished.

The AGN fraction determined by {\sc cigale} is always consistent with zero,
except for the galaxy J123633.2$+$620834.3, for which a 20\% AGN contribution
to the FIR dust luminosity is suggested. Most of the IR excess due to a
potential AGN should be in the MIR, while the SPIRE flux is expected to be
dominated by the starburst component \citep{hatziminaoglou10}; {\sc cigale}
accounts for both.
   
The vast majority of the dropout
galaxies are not detected by SPIRE with the present detection limits.
If LBGs were to follow \cite{meurer99} relation (relating UV-attenuation
to dust emission), what fraction of them would
be detectable? We can use the $(U - V)$ color as a proxy for the rest-frame
$m[{\rm FUV}] - m[{\rm NUV}]$. Simulations suggest that this color provides
$\beta$ with an accuracy better than 0.01 for power-laws
$f_\lambda \propto \lambda^\beta$ and $\beta = -2$, $-1$ and 0. In detail, we can estimate:
\begin{eqnarray}
\beta &=& \log [f_{\nu} ({\rm FUV}) / f_{\nu} ({\rm NUV})] /
 \log [\lambda_{\rm FUV} / \lambda_{\rm NUV}] - 2.00\nonumber\\
 &=& 5.39 \log [S_{\nu} ({\rm FUV}) / S_{\nu} ({\rm NUV})]  - 2.00.
\end{eqnarray}

From the $\beta$ values, we estimate $\log (L_{\rm IR}/L_{\rm FUV}$) for
the LBGs and, after evaluating $L_{\rm FUV}$ from the filter closest to
$\lambda_{\rm FUV}$=0.15\,nm (as a function of the redshift), we
determine  $L_{\rm IR}$. These values of $L_{\rm IR}$ can be
transformed into $L_{\rm 250}$ and $S_{250}$ using the following calibrations
computed from the \cite{dale02} models with $1.0\,{\leq}\,\alpha\,{\leq}\,2.5$,
i.e., star-forming galaxies with similar properties to LBGs.
We have also checked that \cite{chary01} models are consistent with our
calibration.  By fitting polynomials as a function of redshift we find:

\begin{equation}
\log L_{\rm IR} = \sum_{i=0}^3 C_i z^i
 + \log(\nu L_\nu)\left|_{\lambda}\right. ,
\end{equation}
with $C\equiv(C_0,C_1,C_2,C_3)= $
 (1.168, $-$1.166, 0.565, $-$0.091) for $\lambda=250\,\mu$m,
 (1.689, $-$1.426, 0.582, $-$0.089) for $\lambda=350\,\mu$m,
and (2.336, $-$1.605, 0.556, $-$0.078) for $\lambda=500\,\mu$m.

Assuming the \cite{meurer99} relation allows to estimate rough order of magnitude $L_{\rm IR}/L_{\rm FUV}$ for our LBGs \citep{burgarella09}. At $250\,\mu$m we should have
detected 10 LBGs from the sample undetected in the mid-IR and far-IR, three
LBGs from those detected at 24$\,\mu$m, and 2 LBGs from those detected at
250$\,\mu$m.

\begin{figure*}
   \includegraphics [width=130mm, angle=90] {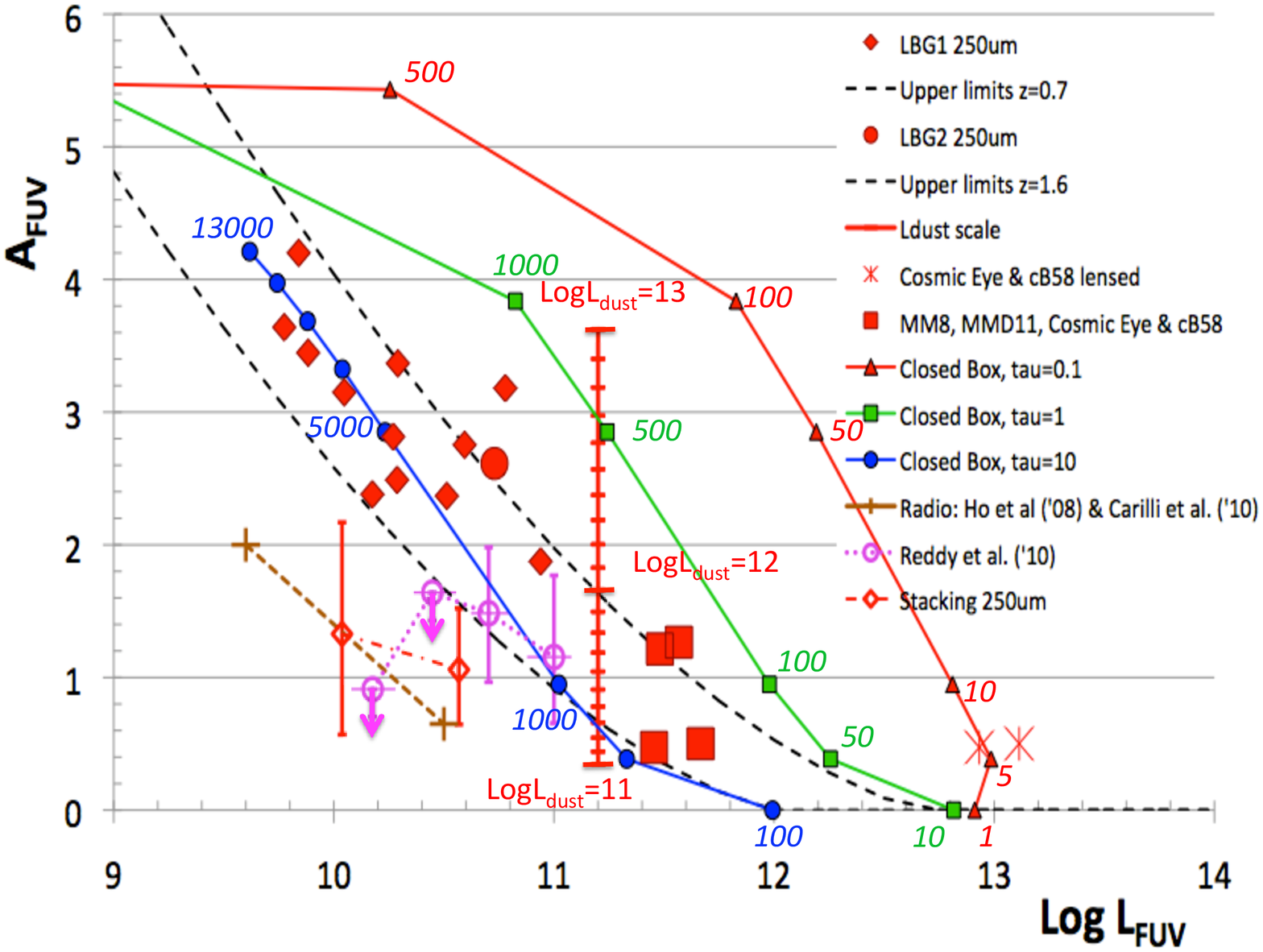}
   \caption{
Red diamonds are LBGs detected by
{\it Herschel}. The two big red stars are the two high redshift LBGs
(the Cosmic Eye and cB58) as observed, while the big red squares are the
same sources after correcting for the amplification plus two unlensed ones. 
The blue dots, green boxes and red triangles are the closed-box models,
plotted as a function of time (age in Myrs) increasing from the bottom-right 
to the top-left part of the diagram. All the models would scale to the left with decreasing mass.
The vertical scale provides the values of $L_{\rm IR}$
in steps of 0.1dex. The two crosses linked by a dashed line are radio-based
measurements. The purple open dots are the stacked points from \cite{reddy10}. 
The red open diamonds corresponds to stacked points at $z\,{\sim}\,1$ from
our sample. The bottom line of this figure is that the
{\it maximum} dust attenuation appears to decline with increasing $L_{dust}$.
} 
   \label{AfuvLfuv}%
 \end{figure*}

For the LBGs not detected by {\it Herschel}, we can estimate upper limits of
$\log (L_{\rm IR}$) for each SPIRE band and we use the lowest of the three values as the 
the final upper limit on $L_{\rm IR}/L_{\rm FUV}$. 
From $\log (L_{\rm IR}/L_{\rm FUV}$) we can estimate $A_{\rm FUV}$.
Fig.~\ref{AfuvLfuv} suggests that the {\it maximum} level of attenuation depends
on $L_{\rm FUV}$ -- the most UV-luminous LBGs yield a lower maximum
$A_{\rm FUV}$ than less UV-luminous ones. This is true for upper limits
as well as for detections. The two higher redshift lensed LBGs
detected in FIR or in sub-mm
(the ``Cosmic Eye'', \citealt{siana09} and cB58, \citealt{siana08}) also
comply with this upper boundary, if we correct for the amplification. 
As do the two unlensed LBGs MM8 \citep{chapman00} and MMD11 \citep{chapman09}. 
We stress that most of the upper limits should populate the area below the
observational limits but none would prevent us from detecting LBGs that would
have larger FUV dust attenuations than the one suggested by the present data 
for a given $L_{\rm FUV}$.

\cite{reddy10} suggest that objects with a lower UV luminosity 
at $1.5\,{\leq}\,z\,{\leq}\,2.6$ have lower bolometric luminosities than UV-bright 
galaxies which, in turn, may suggest lower dust attenuations if we account
for the relation between bolometric luminosities and dust attenuations. This is in agreement with \cite{bouwens10} at much higher redshifts. But, 
when computing A$_{FUV}$ from the stacked points of \cite{reddy10}, we find a 
trend similar to ours for most of the points. \cite{carilli08} and \cite{ho10} estimate the UV dust attenuation by comparing
radio-based star formation rates to UV-based ones using a stacking analysis.
We show Fig.~\ref{AfuvLfuv}, the two points corresponding to
different $\log L_{\rm FUV}$ and they agree with the above trend. \cite{burgarella06} reached a similar conclusion
using {\it Spitzer\/} $24\,\mu$m observations of a sample of LBG
at $z\,{\sim}\,1$. Finally, we have divided our $z \sim  1$ sample in two sub-samples as a function of L$_{FUV}$
and stacked them in the 250$\mu m$ image (Fig.~\ref{AfuvLfuv}).
We find the same trend, again. Note, however, that accounting for the error bars, our stacking and \cite{reddy10},'s are both consistent with a constant $A_{\rm FUV}$. Also, whenever the selection is not fully complete, one may miss objects in the regions where the brightest sources (UV and IR) lie. This effect due to the inhomogeneous background produces holes in the stacking (see e.g. \citealt{bavouzet08}). For this very reason, the z $\sim$ 1 stacked points are only considered as lower limits and we were not able to stack z $\sim$ 2 LBGs.

To understand the
origin of this effect, we build a simple closed box model
(see \citealt{pagel97}),
assuming several exponentially decreasing star formation histories
$\Psi(t) = \Psi_0 e^{-t/\tau}$, with $\tau$ = 0.1, 1 and 10\,Gyr. We assume a
mass of cold gas $M_{\rm gas}$ that forms stars following a Salpeter
initial mass function, and thus produce heavy elements.
$M_{\rm gas}$ evolves as follows:
\begin{equation}
dM_{\rm gas} / dt = -\Psi(t) + E(t),
\end{equation}
where $E(t)$ is the mass ejected by stars at the end of their lifetime. 

The oxygen abundance $Z_{\rm O}$ can be estimated as
\begin{equation}
Z_{\rm O} = -p_{\rm O} \ln [1 - \alpha (1 - e^{-t/\tau})],
\end{equation}
where $p_{\rm O}$ is the oxygen yield and $\alpha$ is the fraction of mass
kept in stellar remnants. We estimate an empirical relation from
\cite{reddy10} that links $12+\log({\rm O}/{\rm H})$ to
$L_{\rm IR}$:
\begin{equation}
\log (L_{\rm IR}/L_{\rm FUV}) = 1.67 ~(12+\log({\rm O}/{\rm H})) -12.72.
\end{equation}
\cite{reddy10}'s objects are in the redshift range $1.5\,{\leq}\,z\,{\leq}\,2.6$, so very close to ours and with metallicities in the range 8.2 $<$ 12/log(O/H) $<$ 8.8 which corresponds to ages in the range of a few 100 Myrs to 5 Gyrs. So, strictly speaking, our models are extrapolations for low $\tau$'s but are in the good range for $\tau$ = 10 Gyrs. Finally, from $L_{\rm IR}$ we compute $A_{\rm FUV}$ using the relationship from \cite{burgarella05}.

Fig.~\ref{AfuvLfuv} shows that this simple closed box
model follows the same trend as our LBGs in the diagram. The initial mass of
gas is set to $\log (M_{\rm gas}/{\rm M}_\odot) =10.5$ to explain the low
redshift LBGs and $\log (M_{\rm gas}/{\rm M}_\odot)=11.0$ (not plotted) for
the high redshift LBGs. This is in agreement with the mass of cold gas
predicted by models \citep[e.g.,][]{lacey11}. We find that star-formation
timescales of $\tau \la 1$ Gyr seem to be ruled out by this model.  

\section{Conclusions}

We have selected two samples of LBGs at $z\,{\sim}\,1$ and $z\,{\sim}\,2$.
For the first time, we can put constraints on the dust emission and therefore
the dust attenuation of LBGs directly from rest-frame FIR
measurements of individual LBGs observed
with {\it Herschel}-SPIRE. Two main conclusions can be drawn from this
analysis.

We detected 12/260 $\sim$ 4.6\% and 1/1558 $\sim$ 0.06\% of the LBGs
at $0.7\,{\leq}\,z\,{\leq}\,1.6$ and $1.6\,{\leq}\,z\,{\leq}\,2.8$,
respectively.  All the other LBGs are undetected by SPIRE, and their dust
attenuation is lower than the detected LBGs.. However, we
have to account for the fact that the limits depend on $L_{\rm FUV}$.

Secondly, the {\it maximum} dust attenuation in the FUV decreases as UV
luminosities increase. 

Other points of interests are:
\begin{itemize}

\item The dropout selections presented in this paper are very efficient
($\sim$95\%) at detecting galaxies in the redshift range \mbox{\boldmath{$0.7\leq z \leq 2.8$}}. 
\item The rest frame UV SEDs of the two dropout samples are similar to
higher redshift LBGs.
\item {\sc cigale} is able to model the observed SEDs from the FUV to the
radio and we provide the derived physical parameters.
\item The stellar masses of these IR-bright dropout galaxies are of the
same order as the stellar masses of IR-bright LBGs observed in IRAC and
MIPS bands.
\item We propose that all LBGs lie in a triangle-shaped region in the $A_{\rm FUV}$ 
vs. $Log L_{\rm FUV}=0$ diagram limited by dust-free (small and/or young) galaxies to the bottom and by the locus for evolving most massive galaxies to the top. 

\end{itemize}

\acknowledgments

SPIRE has been developed by a consortium of institutes led by Cardiff
University (UK) and including Univ. Lethbridge (Canada); NAOC (China);
CEA, OAMP (France); IFSI, Univ. Padua (Italy); IAC (Spain); Stockholm
Observatory (Sweden); Imperial College London, RAL, UCL-MSSL, UKATC, Univ.
Sussex (UK); and Caltech/JPL, IPAC, Univ. Colorado (USA). This development
has been supported by national funding agencies: CSA (Canada); NAOC (China);
CEA, CNES, CNRS (France); ASI (Italy); MCINN (Spain); Stockholm Observatory
(Sweden); STFC (UK); and NASA (USA). The data presented in this paper will
be released through the {\it Herschel\/} Database in Marseille HeDaM
(http://hedam.oamp.fr/HerMES) This work makes use of TOPCAT
http://www.star.bristol.ac.uk/~mbt/topcat/. 

%{\it Facilities:} \facility{Herschel}, \facility{HST}, \facility{Kitt Peak}, \facility{Subaru}.

\clearpage

\end{document}